\documentclass[
apl,
aps,
twocolumn,
superscriptaddress,
amsmath,
amssymb,
floatfix,
]{revtex4-2}

\usepackage{graphicx} % Required for inserting images
\usepackage{xcolor}

\begin{document}

\title{%Low-temperature
Non-Fermi liquid behavior in La$_3$Ni$_2$O$_7$ thin films under hydrostatic pressure}

\author{Deepak Kumar}
\author{Jared Z. Dans}
\author{Keenan E. Avers}
    \affiliation{Maryland Quantum Materials Center and Department of Physics, University of Maryland, College Park, Maryland 20742, USA}
    
\author{Ryan Paxson}
\author{Ichiro Takeuchi}
    \affiliation{Maryland Quantum Materials Center and Department of Physics, University of Maryland, College Park, Maryland 20742, USA}
    \affiliation{Department of Materials Science and Engineering, University of Maryland, College Park, MD, USA}

\author{Johnpierre Paglione}
    \affiliation{Maryland Quantum Materials Center and Department of Physics, University of Maryland, College Park, Maryland 20742, USA}
    \affiliation{Canadian Institute for Advanced Research, Toronto, Ontario M5G 1Z8, Canada}
    \email{paglione@umd.edu}
        \thanks{Authors to whom correspondence should be addressed.}

%\date{December 2025}
\begin{abstract}
The discovery of superconductivity in bilayer nickel-oxides has revived an intense effort to understand the potential of high-temperature superconductivity in these materials and their relation to cuprate superconductors. 
In this work, we investigate the growth and properties of bilayer La$_3$Ni$_2$O$_7$ thin films as a function of substrate, oxygen treatment and applied pressure in order to study the evolution of transport properties. We report epitaxial growth of La$_3$Ni$_2$O$_7$ thin films on LaAlO$_3$ (LAO) (001) and SrLaAlO$_4$ (SLAO) (001) substrates, and the effects of \textit{ex-situ} annealing in a high-pressure furnace under an oxygen-rich environment (15 bar). 
Transport measurements show that the La$_3$Ni$_2$O$_7$ thin films on LAO(001) exhibit Fermi liquid-like metallic behavior with a slight Kondo-like upturn at low temperatures, 
%namely \textit{$\rho$} $\sim$ {-ln(\textit{T}) + \textit{T$^{\gamma}$}}, 
which evolves with the application of modest hydrostatic pressures
toward non-Fermi liquid behavior with a temperature dependence of resistance approaching $\sim T^{1.4}$ at 1.41 GPa.
%When the pressure is increased to 1.41 GPa, the temperature dependence of resistance conforms to a simple power law: \textit{R(T)} = \textit{R$0$} + \textit{AT$^{\alpha}$}, with the exponent ${\alpha}$ approaching 1.4, indicating an onset of non-Fermi liquid behavior. 
The ability to tune the normal state resistivity of La$_3$Ni$_2$O$_7$ films to display non-Fermi liquid behavior under such a modest hydrostatic pressure range -- only 6 - 8 \% of that typically applied via diamond anvil cell (DAC) in La$_3$Ni$_2$O$_7$ single crystals to achieve comparable effects -- is both noteworthy and unexpected. These findings imply the 
strong tunability of La$_3$Ni$_2$O$_7$ in thin film form and the likely proximity of a strongly fluctuating ordered state leading to non-Fermi liquid behavior under even modest applied pressures.

\end{abstract}

\maketitle

\section{Introduction}
The discovery of superconductivity in the infinite-layer nickelates in 2019 \cite{Li} has sparked significant interest across the scientific community, leading to a surge of dedicated research efforts. The emergence of superconductivity in nickelates materials (Ni$^{1+}$, \textit{3d$^9$} electronic configuration), with electronic structures akin to the extensively studied cuprates (Cu$^{2+}$, \textit{3d$^9$} electronic configuration), had been anticipated for several decades but had not been previously observed until recently. This advancement represents a pivotal milestone for the field and introduces both new opportunities and challenges for elucidating the mechanisms underlying high transition temperature (\textit{T$_c$}) superconductivity.
Among these compounds, the Ruddlesden-Popper (RP) bilayer nickelate La$_3$Ni$_2$O$_7$ stands out as the most recently identified high-\textit{T$_c$} superconductor, exhibiting an onset \textit{T$_c$} near 80~K in bulk crystals subjected to 14 GPa of external pressure \cite{Sun}. The presence of structural polymorphs, including various RP variants, presents substantial obstacles to achieving superconductivity in nickelates, often impeding the realization of a zero-resistance state \cite{Chen}. Subsequent studies have demonstrated that partial substitution of La with Pr, or the introduction of small amounts of Sr, mitigates the formation of inter-growth phases, enhances the bilayer RP structure conducive to superconductivity, and increases the superconducting volume fraction \cite{Wang}. Nevertheless, the necessity of applying high pressure to induce superconductivity in La$_3$Ni$_2$O$_7$ crystals remained a considerable limitation for experimental investigations into the underlying mechanisms.\\
Recent advancements have addressed this challenge, with the observation of ambient-pressure superconductivity in La$_3$Ni$_2$O$_7$, (La,Pr)$_3$Ni$_2$O$_7$, La$_{3-x}$Sr$_x$Ni$_2$O$_7$ and (La,Pr,Sm)$_3$Ni$_2$O$_7$ thin films grown under compressive strain on SrLaAlO$_4$ (001) substrates, seemingly eliminating the need for high-pressure conditions in these systems \cite{Hwang1, Hwang2, Hao, Haoran}.
Superconductivity in these films has so far only been observed in those annealed with ozone or grown in an ozone-rich environment, while the pristine, untreated films tend to be either insulating or metallic. Additionally, even pristine films that do not exhibit superconductivity under normal conditions can display such superconducting properties when subjected to high pressure \cite{Osada1}. This makes it essential to investigate the specific roles of O$_2$/O$_3$ and external pressure in bilayer RP thin films. In this work, we present our study on La$_3$Ni$_2$O$_7$ thin films prepared under various strain conditions, emphasizing the role of ozone in achieving superconductivity in nickelate films and demonstrating how applied pressure influences their low-temperature properties. Specifically, we show that there is a change in the power law behavior of the transport	measurement, resulting in non-Fermi liquid behavior for the La$_3$Ni$_2$O$_7$ thin film under modest hydrostatic pressure.

\section{Experimental}
La$_3$Ni$_2$O$_7$ thin films were fabricated on LaAlO$_3$ (001), SrLaAlO$_4$ (001), and YAlO$_3$ (110) substrates by using pulsed laser deposition. Ablation of polycrystalline La$_3$Ni$_2$O$_7$ and SrTiO$_3$ targets was performed using a KrF excimer laser (248 nm). The SrTiO$_3$ here served as the capping layer to La$_3$Ni$_2$O$_7$ film. The laser frequency was set to 5 Hz for La$_3$Ni$_2$O$_7$ and 2 Hz for SrTiO$_3$, with corresponding fluences of 0.5 J/cm$^2$ and 0.6 J/cm$^2$. Prior to deposition, LaAlO$_3$ and YAlO$_3$ substrates were annealed at 800 $^{\circ}$C under 10$^{-5}$ torr of O$_2$ for 30 minutes, while SrLaAlO$_4$ substrates were treated following recipe in ref. \cite{Biswas}, to achieve atomically flat surfaces. Substrate temperatures were maintained at 750 $^{\circ}$C for the La$_3$Ni$_2$O$_7$ layer and 650 $^{\circ}$C for the SrTiO$_3$ layer. Oxygen partial pressures during growth were adjusted to 100 mtorr for La$_3$Ni$_2$O$_7$ and 150 mtorr for SrTiO$_3$, with a fixed target-to-substrate distance of 5 cm. Post-deposition, the films were annealed briefly at 650 $^{\circ}$C, 80 torr for 10 minutes. Film characterization was conducted using X-ray diffraction with a Rigaku diffractometer equipped with a monochromated \textit{Cu} K${\alpha}$1 radiation source (${\lambda}$ = 1.5406 ${\AA}$). Temperature-dependent resistance measured via the four-probe method, and Hall effect measurements under ambient pressure were performed using a Physical Property Measurement System (PPMS, Quantum Design). For high pressure oxygen treatment, samples were put into ceramic crucibles, placed within a quartz tube in a GSL-1100X-RC high-temperature, high-pressure furnace, and annealed for 18 hours at an oxygen pressure of 15 bar.
High-pressure measurements (up to 1.41 GPa) were carried out using a piston-cylinder cell (PCC) with Daphne 7575 oil as the pressure-transmitting medium. The pressure was determined \textit{in-situ} with a \textit{Pb} manometer.
\section{Results and discussion}

\begin{figure}
\centering
\includegraphics[width = 0.4\textwidth]{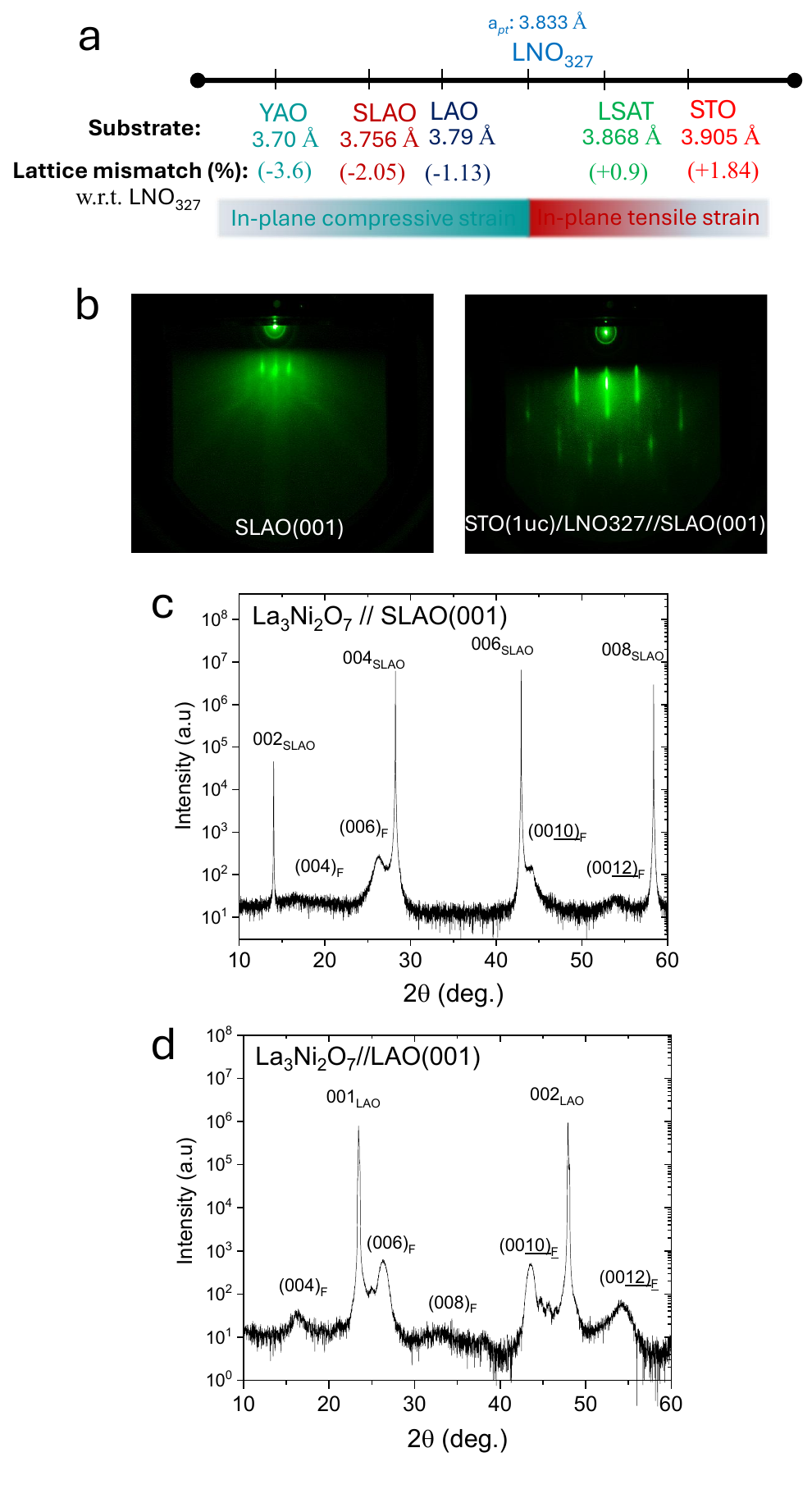}
\caption{\textit{(a) Schematic representation of lattice mismatch between LNO$_{327}$ and various oxide substrates. (b) Reflection High Energy Electron Diffraction (RHEED) patterns for the bare SLAO(001) substrate (left panel) and that of the film hetrostructure (right panel). (c),(d) X-ray Diffraction (XRD) $\theta$-2$\theta$ scans for LNO$_{327}$ films on SLAO(001) and LAO(001) substrates, respectively. The (00\textit{l})$_F$ diffraction peaks belong to the LNO$_{327}$ thin film.}}
\label{fig 1}
\end{figure}

Figure 1 presents an overview of the structural characterization of 6-7~nm La$_3$Ni$_2$O$_7$ (LNO$_{327}$) thin films deposited on LaAlO$_3$ (LAO)(001) and SrLaAlO$_4$ (SLAO)(001) substrates, which exhibit nominal lattice mismatches of approximately -1.1 \% and -2 \%, respectively, as depicted schematically in Figure 1a. The negative values indicate that the films experience compressive strain during growth. The film growth was carefully monitored via \textit{in-situ} reflection high energy electron diffraction (RHEED). Figure 1b illustrates RHEED images of the SLAO(001) substrate (left) and the STO/LNO$_{327}$ film on SLAO (right). 
\begin{figure*}
\centering
\includegraphics[width = 0.8\textwidth]{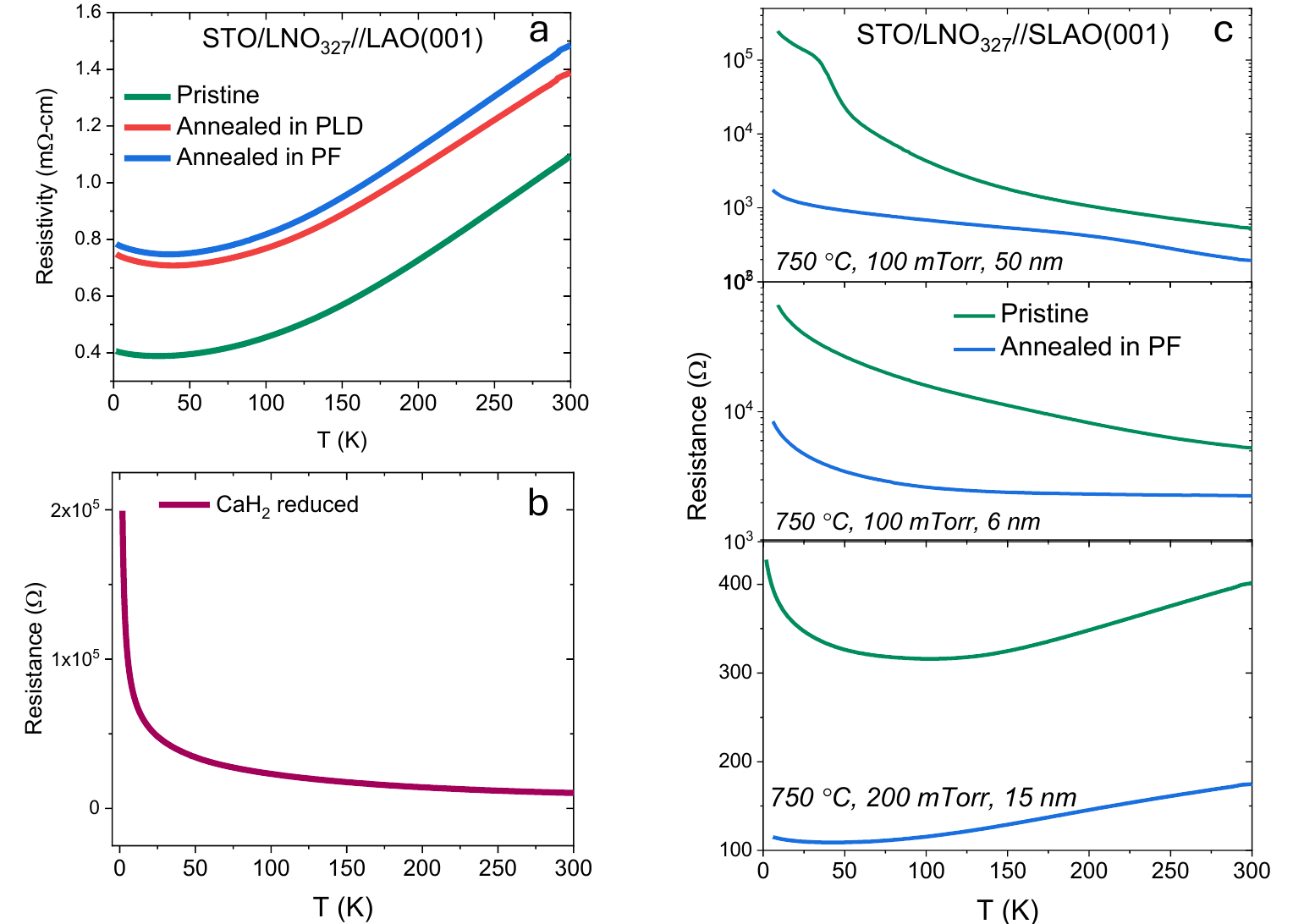}
\caption{\textit{(a) Resistivity vs. temperature plots for the as-grown LNO$_{327}$ film on LAO(001) and the corresponding scans for the annealed films in \textit{in-situ} PLD (at 300 torr, 650 $^{\circ}$C) and in high pressure furnace (PF) (at 15 bar, 650 $^{\circ}$C) for 18 hrs. Temperature dependence of resistance for (b) topotactically reduced LNO$_{327}$ thin film on LAO substrate, (c) pristine and annealed LNO$_{327}$ films on the SLAO(001) substrate. Annealing of films on SLAO(001) substrate was performed in presssure furnace at 15 bar, 650 $^{\circ}$C for 18 hrs.}}
\label{fig 2}
\end{figure*}
A single-unit-cell STO capping layer was deposited to protect the underlying LNO$_{327}$ from degradation during subsequent experimental procedures. Well-defined Kikuchi lines observed on the bare substrate confirm its single-crystalline nature and surface smoothness. Furthermore, the streak-like RHEED pattern recorded for the film in both zeroth and first order diffraction (Figure 1b, right) attest to the epitaxial growth of the film.
Figures 1c and 1d display the X-ray diffraction (XRD) $\theta$ - 2$\theta$ symmetric scans for LNO$_{327}$ films on SLAO and LAO substrates, respectively. Distinct (00\textit{l}) diffraction peaks associated with the bilayer LNO$_{327}$ (n = 2 in Ln$_{n+1}$Ni$_n$O$_{3n+1}$) Ruddlesden-Popper structure are evident, notably the (008) and (0012) reflection, whose intensity has previously been linked to the superconductivity \cite{Hwang2}. The extracted \textit{out-of-plane} $c$-axis lattice constant for the LNO$_{327}$ film was determined to be 20.35 {\AA} for the SLAO substrate and 20.25 {\AA} for the LAO substrate. These values are consistent with prior studies of superconducting LNO$_{327}$ films, indicating that the films are fully strained with the substrates, with more compressively-mismatched SLAO substrate producing longer \textit{out-of-plane} lattice constant. The synthesis of LNO$_{327}$ films is highly sensitive to the kinetic parameters and ambient oxygen partial pressure during deposition, necessitating operation within a narrowly defined growth window to achieve the desired phase formation. 

Figure 2 presents the transport measurements of LNO$_{327}$ films on LAO(001) (Figures 2a and 2b) and SLAO(001) (Figure 2c) substrates. The film grown on an LAO substrate demonstrates typical Fermi-liquid metallic behavior, consistent with previous studies \cite{Hwang1, Gupta}. Since nickelate thin films are often affected by oxygen deficiencies \cite{Hwang2, Hwang1, Lu, Zhou}, additional annealing was conducted both \textit{in-situ} in the PLD chamber and in a furnace under high oxygen pressure to replenish possible oxygen deficiency. Following these treatments, the resistivity of the LNO$_{327}$ film on LAO substrate doubled; however, its resistance-temperature characteristics remained unaltered. The observed increase in resistance may also result from changes in contact resistance subsequent to the oxygen annealing. 

\begin{figure*}
\centering
\includegraphics[width = 0.8\textwidth]{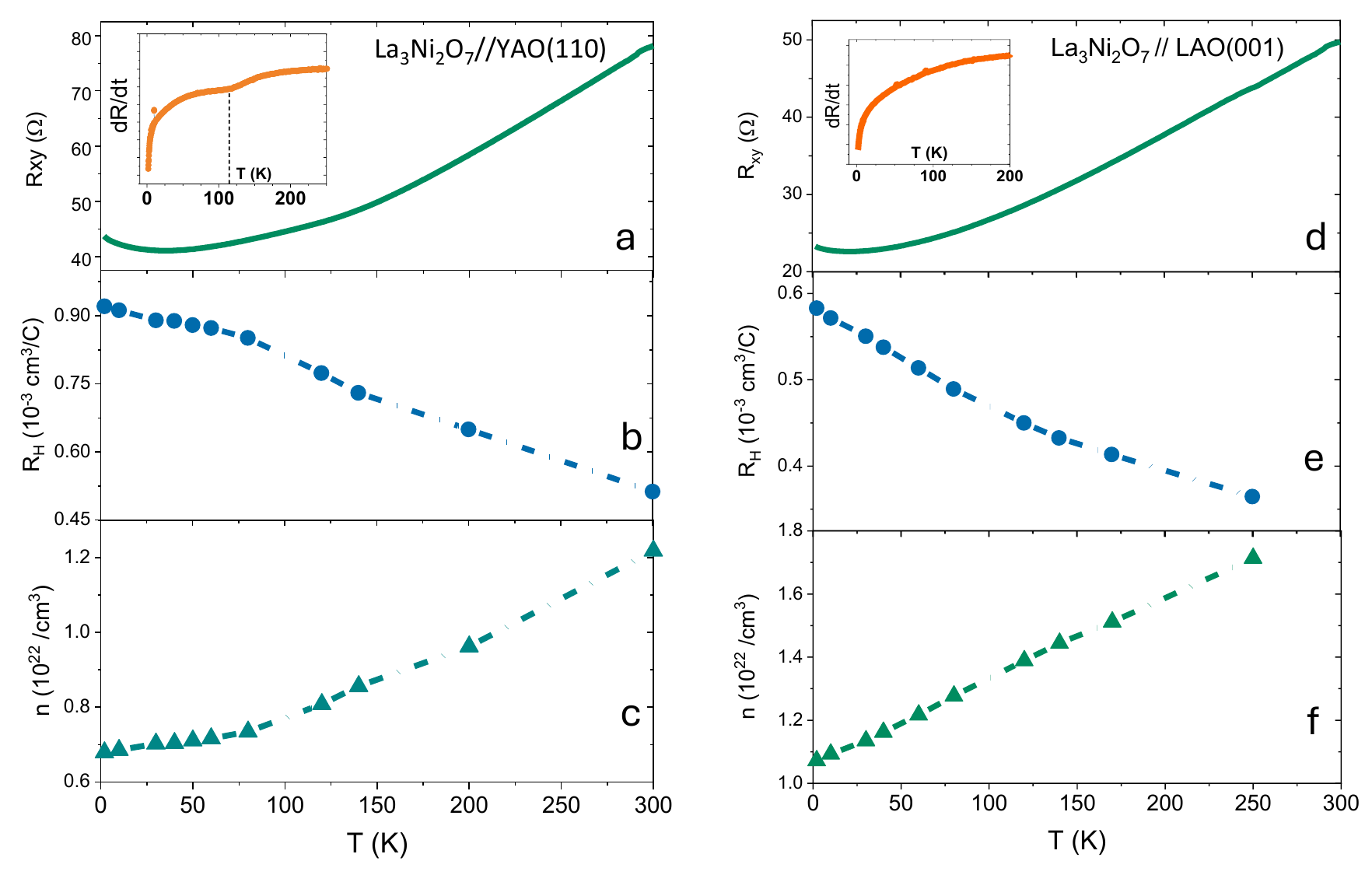}
\caption{\textit{Hall effect measurements for LNO$_{327}$ films on the YAO(110) and LAO(001) substrates; (a) Sheet resistance (b) Hall coefficient (R$_H$), (c) carrier concentration (n) plotted as a function of temperature for the LNO$_{327}$ film on YAO(110) substrate.(d) Sheet resistance (e) Hall coefficient (f) carrier concentration plotted with temperature for LNO$_{327}$ film on the LAO(001) substrate. The dashed line shows transition around 117 K, and is only guide to the eyes.}}
\label{fig 3a}
\end{figure*}
To further investigate the effects of reduction, the sample was topotactically reduced by sealing it in a Pyrex glass tube with approximately 0.1 g of CaH$_2$ powder and heating at 300 $^{\circ}$C for two hours. This procedure parallels the reduction used to obtain the infinite-layer nickelate with superconducting properties (\textit{T$_c$} $\sim$ 15–20~K) from the standard perovskite 113 phase \cite{Li, Osada}. After the reduction, the film turns insulating as anticipated for oxygen-deficient LNO$_{327}$ films \cite{Osada1}. Figure 2c displays four point probe resistivity for pristine and annealed LNO$_{327}$ films on SLAO(001) substrates, grown under varying oxygen pressures and film thicknesses. The pristine films grown at lower oxygen pressure show insulating behavior (Fig. 2c, top and middle panels), whereas the film grown at higher pressure exhibits a metal-to-insulator transition (Fig. 2c, lower panel). This finding supports previous reports \cite{Hwang1, Hwang2} and confirms that a precise growth partial pressure—specifically around 100 mtorr—is required to produce pure bilayer LNO$_{327}$ films (see Fig. 1c and Fig. 1d). If the oxygen partial pressure is higher, the competing LaNiO$_3$ phase forms instead, which exhibits metallic properties in normal state transport. Following the annealing process, films grown at lower pressures continue to exhibit insulating behavior (Fig. 2c, top and middle panels), whereas the film synthesized at higher pressure demonstrates increased metallicity (Fig. 2c, lower panel). Notably, despite annealing under elevated oxygen pressures, the films largely retain their insulating nature—particularly those grown at lower partial pressures where improved conductivity would be anticipated considering oxygen vacancies have been filled. This result contrasts with observations in ozone-annealed LNO$_{327}$ thin films on SLAO(001) substrates, which transition to a superconducting state post-ozone annealing process \cite{Hwang1, Hwang2, Hao, Haoran}. The persistence of insulating behavior may stem from incomplete oxygenation of the films, potentially indicating a need for extended annealing durations under high pressure. These findings underscore the importance of carefully regulated ozone exposure for LNO$_{327}$ thin films, a process after which the films start to display the superconducting properties.

Next, we present Hall effect measurements conducted on nickelate thin films deposited on YAlO$_3$ (YAO)(110) and LAO(001) substrates (Fig. 3), using a Van der Pauw wiring configuration and magnetic field applied perpendicular to the sample surface. Figure 3a displays the sheet resistance \textit{R$_{xy}$} as a function of temperature for the film on YAO(110), accompanied by its derivative to emphasize the transition (inset, Fig. 3a). Both the derivative and the raw resistance-temperature data exhibit a pronounced kink near 120 K, suggesting a transition at this temperature. To corroborate this observation, we performed Hall measurements by performing magnetic field dependent Hall resistance of LNO$_{327}$ films measured between 2K and 300 K (not shown here) and extracted the Hall coefficient (\textit{R$_H$}, Fig. 3b) and carrier concentration (\textit{n}, Fig. 3c). As shown in Fig. 3b, the calculated Hall coefficient \textit{R$_H$} is positive at measured temperatures and decreases with increasing temperature. Such behavior may indicate multiband electronic structure of the LNO$_{327}$ films, where electronic structure near the Fermi surface is dominated by Ni ${3d}$ orbitals (3\textit{d$_{z^2}$} and 3\textit{d$_{x^2 - y^2}$}) and is consistent with previous results \cite{Hwang2, Hwang1, Hu}. Notably, both \textit{n} and \textit{R$_H$} deviate from linearity in temperature around the same temperature where a transition is observed in the resistance-temperature measurements. For comparison, analogous measurements were carried out on a film grown on the LAO substrate, which has a lower lattice mismatch; these results demonstrate a typical metallic resistance-temperature curve (Fig. 3d) with no indication of a corresponding transition. Furthermore, Hall effect data for this sample reveal an almost linear dependence of \textit{n} and \textit{R$_H$} on temperature. Similar transitions have been reported for nickelate thin films and single crystals, and were attributed to the spin density wave (SDW) ordering mechanisms \cite{Gupta, Zhao, Khasanov, Xiaoyang, Kaiwen}. Recent studies have demonstrated that LNO$_{327}$ single crystals at ambient pressure exhibit the SDW order below 150 K, as evidenced by the resonant inelastic x-ray scattering and nuclear magnetic resonance measurements \cite{Zhao,Xiaoyang}. This observation can explain the analogous ordering observed in LNO$_{327}$ films grown on YAO(110) substrates (Fig. 3a). Owing to a substantial lattice mismatch of -3.6\%, the film on the YAO substrate remains largely relaxed, thereby closely replicating the behavior of its bulk counterpart. In contrast, the film grown on the LAO substrate is fully strained due to a smaller lattice mismatch, where the SDW state is either significantly weakened or not observable in resistivity measurements. This behavior resembles that seen in the iron-based \cite{Rotter, Engelmann} and heavy-fermion superconductors \cite{Christianson, Kim} where a distinct SDW anomaly is present in the bulk but becomes broadened as a result of epitaxial strain in thin films. Therefore alternative techniques such as magnetic torque or nuclear magnetic resonance (NMR) may be necessary to detect the SDW transition. Additional experiments are warranted to elucidate the nature of the ordering at this transition. \\

\begin{figure}
    \centering
    \includegraphics[width=0.9\linewidth]{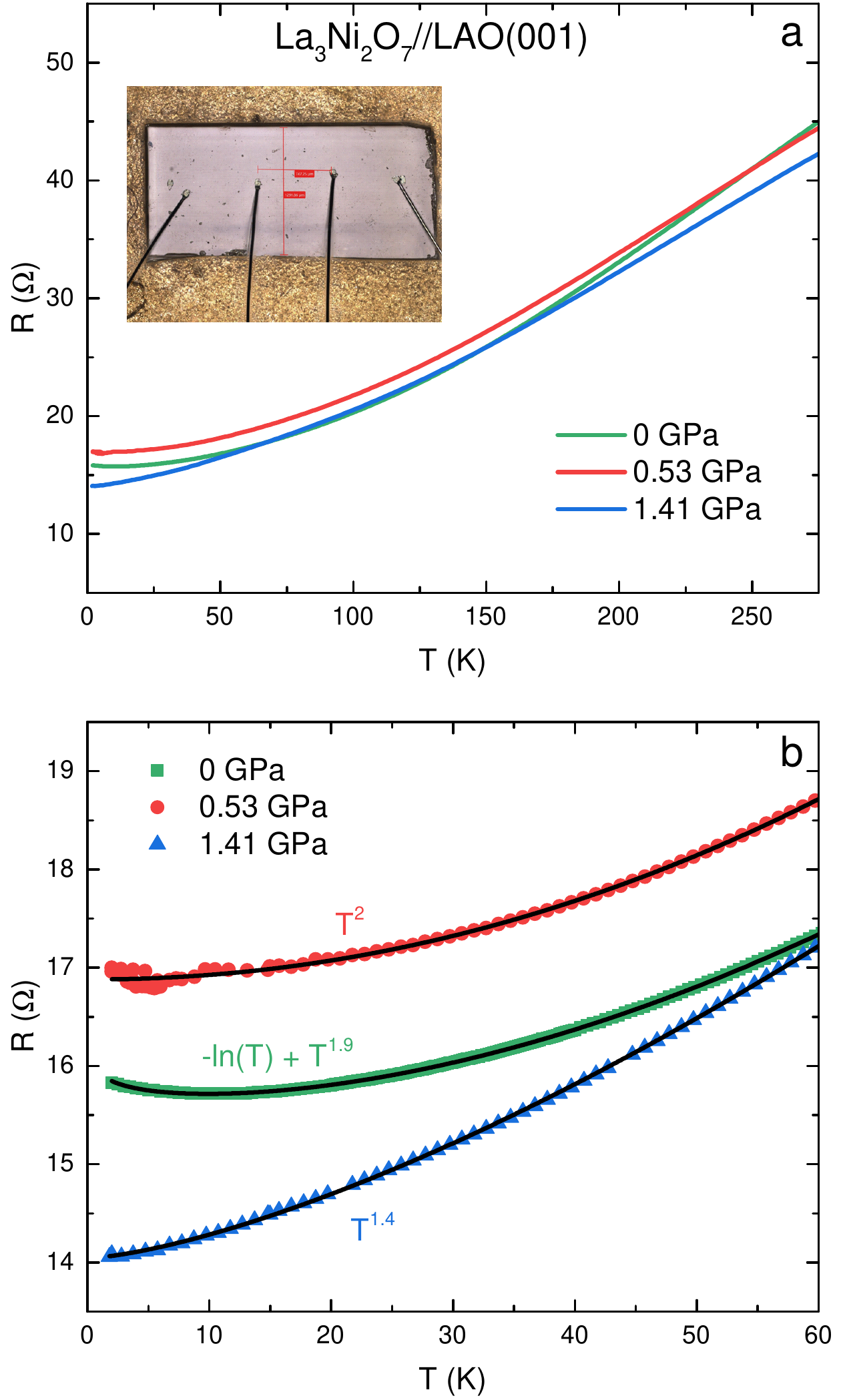}    
    \caption{\textit{(a) Resistance vs temperature scans for the LNO$_{327}$ film on LAO(001) substrate under pressure. (b) Power law fits to the data at low temperature. The symbols represent data points and the black lines represent the fits. The $T$-dependence changes from $T$$^2$ at 0.53 GPa to $T$$^{1.4}$ at 1.41 GPa.}}
    \label{fig 4}
\end{figure}

We now address the pressure dependence of transport measurements in LNO$_{327}$ film grown on LAO(001) substrate. Fig. \ref{fig 4}(a) presents the temperature-dependent resistance \textit{R(T)} at three pressures: ambient, 0.53 GPa, and 1.41 GPa, measured using a piston-cylinder cell (PCC) with Daphne 7575 oil as the pressure-transmitting medium. The pressure was determined \textit{in-situ} with a \textit{Pb} manometer. The LNO$_{327}$ film exhibits metallic behavior across all pressures examined.
The entire \textit{R(T)} curve increases slightly as pressure rises incrementally from ambient pressure to 0.53 GPa, especially below 200 K, which may be attributed to the distortion of the NiO$_6$ octahedra, consistent with previous findings and first-principles investigations \cite{Sun, Yasuhide}, where 1 GPa changes the ground state from metallic to weakly insulating in the LNO$_{327}$ single crystals . 
At ambient pressure, a slight upturn in resistance at low temperature emerges at around 10 K, consistent with prior studies \cite{Hwang1, Ting}, and is potentially attributable to mechanisms such as Kondo-like scattering, oxygen non-stoichiometry, or structural disorder. Notably, this upturn is suppressed under elevated pressures.

To quantify the curvature changes, power-law fitting was performed on the low-temperature data, as shown in Fig. \ref{fig 4}(b). At ambient pressure, the data align best with a Kondo model described by {\textit{R} $\propto$ –ln(\textit{T}) + \textit{T$^{\gamma}$}}. Under increased pressure, the optimal fit transitions to a simple power law: {\textit{R(T)} = \textit{R$_0$} + \textit{AT$^{\alpha}$}}. Up to 0.53 GPa, the measured power law is consistent with Fermi liquid behavior, characterized by a \textit{T}$^2$ dependence of resistance due to quasiparticle-quasiparticle scattering. However, upon further pressure increase this power law evolves to a non-Fermi liquid behavior, with {$\alpha$} = 1.4 over an expansive temperature range spanning from 1.8~K to 60 K. 
%These results indicate that, at near-ambient pressure (0.53 GPa), normal-state transport exhibits Fermi liquid behavior ($\alpha$ = 2). As pressure increases to 1.41 GPa, the exponent $\alpha$ decreases to 1.4 (i.e. \textit{R} $\propto$ \textit{T}$^{1.4}$), reflecting the onset of additional scattering mechanisms and resulting in non-Fermi liquid behavior under moderate hydrostatic pressure.

Comparable non-Fermi liquid transport phenomena have been observed in bulk bilayer nickelates at much higher pressures \cite{Wang, Yanan} along with a theoretical investigation \cite{Onari}, as well as in infinite-layer nickelates \cite{Kyuho}. However, the marked decrease in exponent from $\alpha \simeq$ 2 (Fermi-liquid) to $\alpha=$1.4 with very small relative pressure increase is unexpected, given the very large range of pressures previously required to see substantial changes in this system.  The hydrostatic pressures applied to our thin films amount to only 6 to 8\% of those utilized with Diamond Anvil Cells on LNO$_{327}$ single crystals to achieve similar effects \cite{Sun, Wang, Yanan}. This is very surprising to see such a drastic change in the electronic correlations over a small pressure range, and has not been reported thus far in the LNO$_{327}$ thin films to the best of our knowledge. Given the in-situ tuning of the thin film sample with little relative changes in structure and/or strain, we speculate that this sample lies proximate to a quantum critical point (QCP) of the SDW ordered phase \cite{Gupta}. The behavior observed here is in fact very comparable to that observed in infinite-layer nickelates, Fe-based, and cuprate superconductors \cite{Danfeng, WangN, Duffy, Kontani, Huang}. 
Here, the level of epitaxial strain combined with adequate oxygen-tuning has tuned the LNO$_{327}$ thin film to be very close to the SDW phase at ambient pressure, requiring only a few GPa of applied pressure to tune closest to the QCP. 

For spin fluctuation scattering, the theoretical value of the temperature exponent depends on the nature and dimensionality of the magnetism of the proximate ordered phase, ranging from superlinear in two-dimensional ferromagnetic ($\alpha$=4/3) and three-dimensional antiferromagnetic ($\alpha$=3/2) systems to linear ($\alpha$=1) in 2D AFM systems \cite{Stewart}. 
While measurements over a wider pressure range are required to understand the quantitative ``distance'' to the critical point itself, the rapid change in power law suggests that $\alpha$=1.4 is rapidly approaching the limiting power close to the QCP itself. 
Because a linear exponent has been repeatedly observed in other nickelate systems \cite{Kyuho, Danfeng, Hiragami}, it is likely that this system is expected to approach similar behavior at slightly higher pressures as well, consistent with a scenario where quantum criticality is tuning the unconventional normal state behavior and possibly instigating the superconductivity itself. Further fine tuning of strain, oxygen annealing and pressures will indicate if superconductivity itself can be observed to emerge along with such tuning.

%much like what is seen in cuprate materials near their superconducting optimal doping levels \cite{Ando}.

\section{Conclusions}

In summary, we have achieved the epitaxial growth of La$_3$Ni$_2$O$_7$ thin films on various oxide substrates and subsequently annealed them in an oxygen-rich environment, rather than utilizing ozone. Our findings show that simply annealing the films in oxygen, regardless of its richness, in not enough to induce superconductivity; ozone appears to be necessary for promptly filling oxygen vacancies in the La$_3$Ni$_2$O$_7$ films. Ozone, as a highly reactive form of oxygen, appears to replenish lost oxygen during film growth much more efficiently than the molecular oxygen, which may require several days to achieve comparable compensation. A transition potentially associated with a spin density wave has been identified in Hall measurements of La$_3$Ni$_2$O$_7$ film grown on a YAO(110) substrate. The Hall coefficient remains positive throughout the entire temperature range, but decreases as temperature rises, which may indicate the presence of a multiband electronic structure in La$_3$Ni$_2$O$_7$ films.
Additionally, our results demonstrate that at ambient pressure, La$_3$Ni$_2$O$_7$ films on LaAlO$_3$ (001) exhibit a slight upturn in low-temperature resistivity, which fits well to a logarithmic Kondo model. Under elevated pressure, this upturn vanishes, and the temperature-dependent resistance can be best fitted to a power law of the form {\textit{R(T)} = \textit{R$_0$} + \textit{AT$^{\alpha}$}} with $\alpha$ = 2 (Fermi-liquid) at 0.52 GPa, and $\alpha$ $\sim$ 1.4 (non-Fermi liquid) at 1.41 GPa. These observations underscore the role of quantum critical scattering mechanisms, such as the spin fluctuation scattering, leading to a non-Fermi liquid behavior in La$_3$Ni$_2$O$_7$ thin films. Moreover, it is remarkable that the normal state resistivity of La$_3$Ni$_2$O$_7$ thin films can be tuned to exhibit non-Fermi liquid behavior within a narrow hydrostatic pressure range—specifically, only 6 – 8 \% of the pressure applied with diamond anvil cells in La$_3$Ni$_2$O$_7$ single crystals to observe similar phenomena.

\section{acknowledgments}
We thank Richard L. Greene, T. Sarkar and Shanta R. Saha for discussions and assistance.
This work was supported by the Breakthrough Energy Fellows program, 
the Gordon and Betty Moore Foundation’s EPiQS Initiative Grant No. GBMF9071,
the U.S. National Science Foundation Grant No. DMR2303090%{(sample preparation)}, 
, the Air Force Office of Scientific Research Grant No. FA9950-22-1-0023, 
the Department of Energy, Office of Basic Energy Sciences Award No. DE-SC-0019154%(experimental measurements),
, the National Science Foundation DMREF program (DMR2523217),
the NIST Center for Neutron Research, and the Maryland Quantum Materials Center.

\end{document}